\documentclass[11pt]{article}

\usepackage[final]{acl}

\usepackage{times}
\usepackage{latexsym}

\usepackage[T1]{fontenc}

\usepackage[utf8]{inputenc}

\usepackage{microtype}

\usepackage{inconsolata}

\usepackage{graphicx}

\usepackage{multirow}
\usepackage{amsmath}
\usepackage{array}

\newcolumntype{C}[1]{>{\centering\arraybackslash}p{#1}}

\title{Adaptive Test-Time Inference for Text2Cypher with Trace Budgeting and Selective Refinement}

\author{Makbule Gulcin Ozsoy \\
  Neo4j, London, UK \\
  \texttt{makbule.ozsoy@neo4j.com} \\
}

\begin{document}
\maketitle
\begin{abstract}
Large language models have enabled natural language interfaces for structured databases, but generated queries may still contain syntactic errors, violate database schemas, or fail during execution. Test-time inference strategies improve generation reliability without additional training, but existing approaches often use fixed inference budgets and uniform refinement strategies, leading to unnecessary computation across questions with different complexity levels. In this work, we investigate adaptive test-time inference for Text2Cypher and introduce two strategies: adaptive trace budgeting, which dynamically adjusts the candidate generation budget based on question difficulty, and selective execution-guided refinement, which applies correction only when additional inference is expected to be beneficial. Experiments on Gemma-2-9B and Qwen-2.5-7B show that adaptive trace budgeting reduces the average generation budget by 30.7\% and wall-clock inference time by 21-25\% while maintaining comparable generation quality. Selective refinement preserves nearly all execution success gains of full refinement, reducing execution success by only 0.2--0.5\% while avoiding unnecessary refinement for simpler questions. Experiments show that a single correction model (Gemma-4) effectively refines outputs from a different model family, suggesting refinement transfers across families. 
\end{abstract}

\section{Introduction} \label{sec:introduction}

Large language models (LLMs) have improved natural language interfaces for structured databases, enabling direct generation of executable queries from natural language questions across tasks such as Text2SQL, Text2SPARQL, and Text2Cypher~\cite{yu2025text2sql,d2025investigating,ozsoy2025text2cypher}. However, generated queries may still contain syntactic errors, violate database schemas, or fail during execution, limiting their reliability in practical applications.

In order to improve generation reliability without retraining the underlying models, recent work has explored test-time inference strategies, including confidence-based candidate selection and inference-time grammar and schema filtering~\cite{wang2023self,brown2024large,jivani2025reliable,tuccio2025grammar}. 
While these approaches improve robustness, they generally allocate the same amount of computation to every input question. In practice, question complexity varies considerably: simple questions often require fewer inference steps, whereas more complex questions may benefit from additional candidate generation or correction. 
Moreover, stricter filtering mechanisms may reject all generated candidates for some inputs, resulting in empty predictions that require additional recovery steps. Applying the same generation and refinement strategy uniformly therefore introduces unnecessary inference cost. These limitations motivate adaptive test-time strategies that allocate computation based on query complexity and apply additional processing only when needed.

\begin{figure*}
    \centering
    \includegraphics[width=0.90\linewidth]{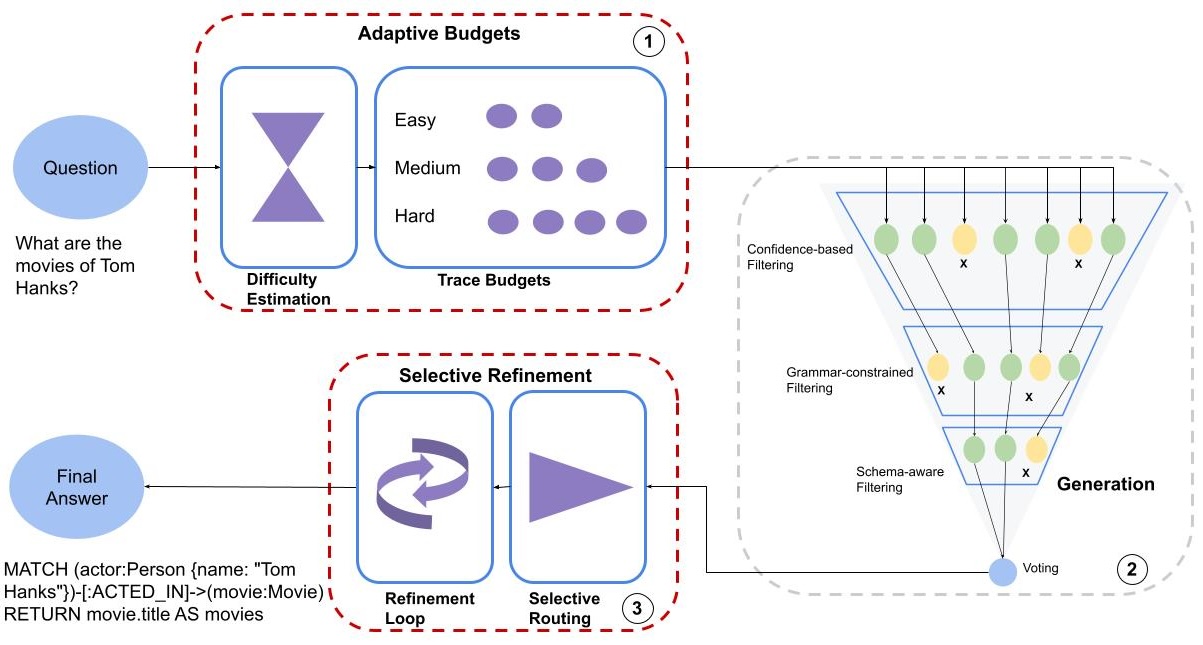}
    \caption{Overview of the proposed framework. Adaptive Trace Budgeting and Selective Refinement (red; this work) extend a confidence-based, grammar/schema-aware generation module (gray) from prior work~\cite{dessi2026, ozsoy2026extending}.
    }
    \label{fig:overview}
\end{figure*}

In this work, we investigate adaptive test-time inference for Text2Cypher, the task of generating Cypher queries for property graph databases from natural language questions. Building on previous confidence-based, grammar- and schema-aware Text2Cypher inference frameworks~\cite{dessi2026,ozsoy2026extending}, we introduce two complementary strategies: (i) adaptive trace budgeting, which estimates question difficulty before generation and dynamically allocates the candidate generation budget, and (ii) selective execution-guided refinement, which applies iterative correction only to predictions likely to benefit from additional inference while avoiding unnecessary refinement for low-complexity queries.  

We evaluate the proposed framework on a Text2Cypher benchmark using two LLMs from different model families, Gemma-2-9B and Qwen-2.5-7B. Adaptive trace budgeting reduces the average generation budget by 30.7\% and wall-clock inference time by 21-25\% while maintaining comparable generation quality. Execution-guided refinement improves execution success, and selective refinement preserves most of this improvement while reducing unnecessary refinement for low-complexity questions. Furthermore, our results suggest that a single refinement model can refine outputs from different generator families, indicating that the generator and correction models do not need to belong to the same model family. 

Our main contributions are as follows:
\begin{itemize}
\item We propose adaptive trace budgeting, allocating candidate generation budgets based on estimated question difficulty to reduce inference cost while maintaining generation quality.
\item We introduce selective execution-guided refinement, applying correction based on question difficulty and filtering outcomes. 
\item We demonstrate the effectiveness of the proposed framework across two LLM families and show that a shared refinement model can generalize across evaluated generator models.
\end{itemize}

The remainder of the paper is organized as follows. Section~\ref{sec:related_work} reviews related work, Section~\ref{sec:methodology} presents the proposed method, Section~\ref{sec:experiments} presents the experiments, and Section~\ref{sec:conclusion} concludes the paper.

\section{Related Work}\label{sec:related_work}

Existing work on Text2Query generation has primarily focused on improving model inputs and training through prompt engineering, schema serialization, context augmentation, grounding, and fine-tuning~\cite{gao2024text,d2025investigating,yu2025text2sql,lee2025safe}. For Text2Cypher, similar approaches include schema-aware prompting and grounding for property graph databases~\cite{mandilara2025decoding,ozsoy2025text2cypher,bunkova2026grounding,cazzaro2026achieving}. While these approaches improve the model or its inputs, more recent work has focused on improving generation quality at inference time.

\paragraph{Test-Time Inference for Query Generation} 
More recently, test-time inference has emerged as an effective alternative for improving generation quality without retraining the underlying language model. These strategies can be broadly grouped into two categories: sampling-based approaches, such as Best-of-$N$ sampling, self-consistency, majority voting, and confidence-based candidate selection or aggregation, which generate multiple candidates and select or aggregate the most reliable one~\cite{wang2023self,brown2024large,huang2025best,fu2025deep,kang2026scalable}; and constrained-decoding approaches, which enforce grammatical or structural constraints during generation~\cite{willard2023efficient,tuccio2025grammar,jivani2025reliable}.

A related line of work investigates adaptive computation, dynamically allocating the sampling budget according to input difficulty or stopping once a reliability criterion is satisfied~\cite{aggarwal2023let,snell2024scaling,damani2025learning}. However, these methods have primarily been studied in general reasoning and text generation rather than query generation. Test-time inference has also been applied to Text2Query generation~\cite{guo2025rethinking,tritto2026gradesql}. For Text2Cypher specifically, confidence-based candidate selection~\cite{dessi2026} has been extended with grammar validation and schema-aware filtering~\cite{ozsoy2026extending}. However, these approaches employ a fixed candidate generation budget and do not adapt inference computation to the difficulty of the input question.

\paragraph{Iterative Refinement for Query Generation} 
Iterative refinement methods have shown that language models can improve their outputs through self-generated feedback and correction~\cite{madaan2023self,shinn2023reflexion,song2025progco}. This idea has also been explored for Text2SQL, where execution feedback drives self-correction or multi-agent repair of faulty queries~\cite{pourreza2023din,chen2024teaching,wang2025mac}. For Text2Cypher, an exploratory study proposed a verify-then-correct loop combining rule-based, execution-based, and LLM-based verification, although without a systematic empirical evaluation~\cite{ozsoy2025iterativerefinement}. More recently, execution-guided refinement has been shown to improve execution success by repairing invalid queries~\cite{jung2026ras,tomczak2026cygnet}. However, these approaches typically apply refinement uniformly to all generated queries, without considering whether additional correction is necessary or beneficial.

To address this limitation, our approach introduces selective execution-guided refinement, which applies refinement only when additional inference is expected to improve the generated query. By combining adaptive trace budgeting with selective refinement, our framework reduces inference cost while maintaining generation quality.

\section{Methodology} \label{sec:methodology}

Our approach builds on previous confidence-based and grammar-/schema-aware Text2Cypher inference frameworks~\cite{dessi2026,ozsoy2026extending}, which generate multiple candidate queries and improve reliability through confidence estimation and structural filtering. While effective, these frameworks apply a fixed inference budget across input questions with different complexity levels. Moreover, strict filtering mechanisms may produce empty outputs when all generated candidates are rejected, motivating additional refinement strategies to improve unreliable predictions. 

Building upon these frameworks, we retain the candidate generation and filtering components and introduce two complementary inference-time strategies: (1) adaptive trace budgeting, which dynamically determines the number of generation traces based on the estimated difficulty of the input question, and (2) selective execution-guided refinement, which applies validation and correction only when additional inference is expected to improve the generated query while avoiding unnecessary refinement for low-complexity questions. Figure~\ref{fig:overview} provides an overview of the proposed framework.

{
\renewcommand{\arraystretch}{1.1} 
\begin{table}
    \centering
    \begin{tabular} {p{0.365\linewidth}p{0.54\linewidth}}
    \hline
    \textbf{Feature Category}    &    \textbf{Examples}                              \\
    \hline
    \hline
    \textbf{Aggregation}            & count, sum, average,  how many, maximum, ... \\
    \textbf{Superlatives}           & most, lowest, best, ...  \\
    \textbf{Comparison}             & more than, less than, at least, above, ...   \\
    \textbf{Temporal}               & before, latest, since, ...  \\
    \textbf{Multi-hop}              & through, connected, path, reachable, ...  \\
    \textbf{Negation}               & not, without, except, never, exclude, ...   \\
    \textbf{Property filtering}     & whose, founded in, with a revenue, located in, ... \\
    \textbf{Question length}        & Length of natural language questions $>=15$   \\
    \textbf{Named entities}         & Quoted entity names   \\
    \hline
    \end{tabular}
    \caption{Heuristic feature categories used for estimating question difficulty.}
    \label{tab:difficulty_features}
\end{table}
}

\subsection{Adaptive Trace Budgeting Strategy}\label{sec:trace_allocation}

Considering question difficulty varies across inputs, we propose an adaptive trace budgeting strategy that dynamically adjusts the generation budget according to the estimated difficulty of each question. 

 Given a natural language question $q$, we estimate the expected query generation difficulty before candidate generation. We employ a lightweight rule-based estimator that extracts linguistic features from the input question. These features capture patterns commonly associated with more challenging Cypher generation, including aggregation requests, superlative expressions, comparison and temporal reasoning, negation, multi-hop traversal cues, property filtering, question length, and named entities. Table~\ref{tab:difficulty_features} summarizes the feature categories used by the estimator. The estimator serves as a lightweight proxy for question difficulty rather than a precise predictor. It requires no additional model or training and adds negligible computational overhead.

Each characteristic is represented as a binary indicator denoting the presence or absence of a complexity cue. Formally, let
\[
F(q)=\{f_1(q),\ldots,f_m(q)\},
\]
where each feature $f_i(q)\in\{0,1\}$ indicates whether the corresponding complexity feature is observed in the input question. The overall difficulty score is computed as
\[
s(q)=\sum_{i=1}^{m}f_i(q).
\] 
The resulting score is mapped to one of three difficulty tiers:
\[
t(q)=
\begin{cases}
\textit{Easy}, & s(q)\le\tau_1,\\
\textit{Medium}, & \tau_1 < s(q)\le\tau_2,\\
\textit{Hard}, & s(q)>\tau_2,
\end{cases}
\]
where $\tau_1$ and $\tau_2$ are difficulty score thresholds. Questions assigned to higher tiers are expected to require more inference computation due to more complex query structures. 

Finally, the estimated difficulty tier determines the candidate generation budget:
\[
N(q)=B(t(q)),
\]
where $N(q)$ denotes the number of candidate traces generated for question $q$, and $B(\cdot)$ is the predefined budget function. In this work, a trace denotes one independently generated candidate Cypher query produced during the candidate generation process. 
Easy questions receive a reduced trace budget, medium questions receive an intermediate budget, and hard questions receive the full budget of the baseline confidence-based inference framework. Since difficulty estimation is performed before generation using deterministic heuristics, its overhead is negligible compared with LLM inference, requiring approximately 0.07\,ms per question.

\subsection{Selective Refinement Strategy}\label{sec:refinement}

\begin{figure}
    \centering
    \includegraphics[width=0.99\linewidth]{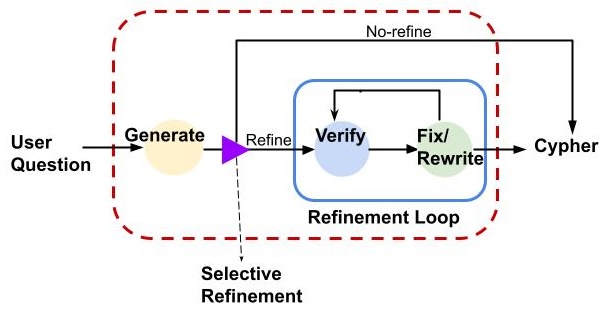}
    \caption{Selective refinement loop: Execution-guided query correction applied selectively based on question difficulty and empty outputs.}
    \label{fig:refinement}
\end{figure}

Refinement iteratively improves an initial query by using feedback from a verifier to guide subsequent corrections. As illustrated in Figure~\ref{fig:refinement}, an initial query is first verified. If verification succeeds, it is accepted. Otherwise, the verifier feedback is provided to a correction LLM, which generates an updated query. Formally, at refinement iteration $i$, the updated query is generated as
\[
c_{i+1}=R(c_i, q, S, e_i),
\]
where $q$ is the input question, $S$ is the database schema, $c_i$ is the current Cypher query, and $e_i$ denotes the feedback returned by the verifier for query $c_i$. The refinement process terminates once the verifier accepts the query or the maximum refinement budget is exhausted.

In this work, we employ execution-guided refinement, following recent work~\cite{jung2026ras}, where query execution on the target database serves as the verifier. Whenever execution returns an error, the input question, database schema, generated Cypher query, and the corresponding execution feedback are provided to a correction LLM to generate a revised query. The correction model may either be the same model used for query generation (self-refinement) or a different language model.

While execution-guided refinement improves generation reliability, applying it uniformly introduces unnecessary inference cost for questions whose initial predictions are already reliable. We therefore propose a selective execution-guided refinement strategy that applies the refinement loop only to questions expected to benefit from additional inference. Specifically, we evaluate refinement policies operating on different subsets of predictions, including all predictions, only empty outputs, and difficulty-based subsets. Using the difficulty tiers introduced in Section~\ref{sec:trace_allocation}, refinement is selectively applied to non-easy questions together with empty outputs produced by the filtering stage. This adaptive strategy retains most of the benefits of uniform refinement while reducing unnecessary refinement for low-complexity questions.

\section{Experiments} \label{sec:experiments}

\subsection{Experimental Setup and Evaluation Metrics}

We follow the experimental setup of previous confidence-based and grammar-/schema-aware Text2Cypher inference studies~\cite{dessi2026,ozsoy2026extending} to enable direct comparison. Experiments are conducted on the same subset of the publicly available Text2Cypher dataset~\cite{ozsoy2025text2cypher}, corresponding to the challenging evaluation split used in previous work~\cite{dessi2026}. This subset contains 789 test questions spanning three graph databases: ``recommendations'', ``companies'', and ``neoflix''.

We evaluate the proposed approach using two language models from different model families: Gemma-2-9B\footnote{\url{https://huggingface.co/google/gemma-2-9b-it}} and Qwen-2.5-7B\footnote{\url{https://huggingface.co/Qwen/Qwen2.5-7B-Instruct}}. Candidate generation is performed using vLLM~\cite{kwon2023efficient} on an NVIDIA A40 GPU environment. Following prior work, generated outputs are post-processed to remove formatting artifacts, such as redundant ``cypher:'' prefixes, before evaluation. For execution-guided refinement, we use the Gemma-4 model from the Ollama library (\textasciitilde8B parameters, Q4\_K\_M quantization)\footnote{\url{https://ollama.com/library/gemma4:latest}, accessed July 2026}. 
Refinement is performed locally using Ollama, with the same correction model applied to both generator models. Reported wall-clock inference times correspond to the complete pipeline under these experimental settings.

Following previous work, we report both translation-based and execution-based evaluation. Translation-based evaluation measures lexical similarity between the generated and reference Cypher queries, while execution-based evaluation executes both queries against the target database and compares their returned results. In addition, we report execution statistics, including execution success rate and error counts, to assess the practical reliability of generated queries. To evaluate adaptive test-time inference, we additionally report computational statistics, including the average number of generated traces and wall-clock inference time. For refinement experiments, we compare refinement policies that apply correction to different subsets of predictions and analyze their impact on query quality, execution success, and refinement overhead. 

{
\renewcommand{\arraystretch}{1.1} 
\begin{table}[t]
\centering
\begin{tabular}{p{0.35\linewidth}C{0.245\linewidth}C{0.28\linewidth}}
\hline
\textbf{Difficulty Tier} & \textbf{Questions} & \textbf{Trace Budget} \\
\hline
\hline
Easy   & 106 (13.4\%) & 21 \\
Medium & 557 (70.6\%) & 30 \\
Hard   & 126 (16.0\%) & 45 \\
\hline
Avg. trace budget & \multicolumn{2}{c}{31.2 (vs. fixed: 45)} \\
Est. reduction & \multicolumn{2}{c}{30.7\%} \\
\hline
\end{tabular}
\caption{Distribution of difficulty tiers and corresponding adaptive trace budgets on the evaluation dataset.}
\label{tab:difficulty_distribution}
\end{table}
}

\subsection{Evaluation Results}

We first analyze the difficulty tier distribution in the evaluation dataset, which guides the adaptive trace budgeting and selective execution-guided refinement strategies. 
Experiments are conducted with Gemma-2-9B and repeated with Qwen-2.5-7B to assess consistency across generator models.

{
\renewcommand{\arraystretch}{1.1} 
\begin{table*}[t]
\centering
\begin{tabular}{lcccc}
\hline
\textbf{Method} &
\textbf{ROUGE-L} &
\textbf{ROUGE-L} &
\textbf{Exec.} &
\textbf{Time} \\
&
\textbf{(Lexical)} &
\textbf{(Execution)} &
\textbf{Success (\%)} &
\textbf{(s)}\\
\hline
\hline
Fixed budget                      & 0.6972 & 0.2929 & 84.0 & 23.4 \\
Adaptive budget$^{\dagger}$                     & 0.6863 & 0.3038 & 83.1 & 17.5 \\
\hline
Fixed b. + refinement (all)$^{*}$              & \textbf{0.7296} & 0.3199 & \textbf{92.6} & 24.5 \\
Adaptive b. + refinement (all)$^{*\dagger}$    & 0.7288 & \textbf{0.3305} & 92.4 & 18.6 \\
\hline
Adaptive b. + refinement (empty+medium+hard)&
0.7279 & 0.3274 & 91.9 & 18.5 \\
Adaptive b. + refinement (empty+hard) &
0.7222 & 0.3145 & 88.3 & 17.8 \\
Adaptive b. + refinement (empty only) &
0.7214 & 0.3127 & 87.8 & 17.6 \\
\hline

\end{tabular}
\caption{
Performance of adaptive trace budgeting and selective refinement on
Gemma-2-9B. 
$^{*}$ indicates statistically significant improvement over the fixed-budget baseline
according to paired bootstrap resampling (10,000 resamples, $p<0.05$). 
$^{\dagger}$ indicates no statistically significant difference from the corresponding
fixed-budget configuration.
}
\label{tab:eval_results}
\end{table*}
}

\subsubsection{Difficulty Tier Distribution} \label{sec:difficulty_tier}

Each question is assigned a difficulty tier using the rule-based estimator described in Section~\ref{sec:trace_allocation}. 
In our experiments, we set thresholds $\tau_1$ and $\tau_2$ based on the difficulty score distribution. Given the mean difficulty score of 1.47, and we set $\tau_1=0$ and $\tau_2=2$, resulting in easy, medium, and hard tiers corresponding to scores of 0, 1--2, and $\geq3$, respectively. The thresholds depend only on the input difficulty scores and were not selected based on evaluation outcomes. Table~\ref{tab:difficulty_distribution} summarizes the resulting distribution on the evaluation dataset, with 106 (13.4\%), 557 (70.6\%), and 126 (16.0\%) questions assigned to the easy, medium, and hard tiers, respectively.
These difficulty tiers are used throughout the remaining experiments to determine adaptive trace budgets and selective refinement policies.

\subsubsection{Effect of Adaptive Trace Budgeting}

We first evaluate whether the proposed adaptive trace budgeting strategy can reduce inference cost while maintaining generation quality. Following the tier-based budgeting described in Section~\ref{sec:difficulty_tier}, easy, medium, and hard questions receive trace budgets of 21, 30, and 45 traces, respectively. Under this setup, the average trace budget decreases from 45.0 to 31.2 traces per question, corresponding to a 30.7\% reduction in generated traces.

Table~\ref{tab:eval_results} compares the proposed adaptive strategy with the best-performing inference configuration from~\citet{ozsoy2026extending}, which combines online confidence-based, grammar- and schema-aware filtering.  
The comparison isolates the effect of replacing the fixed trace budget with the proposed adaptive budgeting. Despite using fewer traces, the adaptive strategy achieves comparable performance. The execution success rate decreases by less than one percentage point (84.0\% vs.~83.1\%), while both lexical and execution ROUGE-L scores remain comparable. The average inference time decreases from 23.4\,s to 17.5\,s per question, corresponding to an approximately 25\% reduction in total inference time. Paired bootstrap resampling (10,000 resamples) confirms that the difference between the adaptive and fixed-budget configurations is not statistically significant.

These results indicate that reducing trace budgets for easier questions while maintaining larger budgets for difficult ones reduces inference cost without substantially affecting generation quality.

{
\renewcommand{\arraystretch}{1.1}
\begin{table*}[t]
\centering
\begin{tabular}{p{0.13\linewidth}p{0.34\linewidth}cccc}
\hline
\textbf{Model} &
\textbf{Method} &
\textbf{ROUGE-L} &
\textbf{ROUGE-L} &
\textbf{Exec.} &
\textbf{Time} \\
&
&
\textbf{(Lexical)} &
\textbf{(Execution)} &
\textbf{Success (\%)} &
\textbf{(s)}\\
\hline
\hline

\multirow{4}{*}{Gemma-2-9B}
& Fixed budget + refinement (all)
& 0.7296 & 0.3199 & 92.6 & 24.5 \\
& Adaptive budget + refinement (all)
& 0.7288 & 0.3305 & 92.4 & 18.6 \\
& Adaptive budget + \newline refinement (empty+medium+hard)
& 0.7279 & 0.3274 & 91.9 & 18.5 \\

\hline

\multirow{4}{*}{Qwen-2.5-7B}
& Fixed budget + refinement (all)
& 0.7261 & 0.2769 & 92.1 & 17.7 \\
& Adaptive budget + refinement (all)
& 0.7263 & 0.2793 & 91.6 & 14.0 \\
& Adaptive budget + \newline refinement (empty+medium+hard)
& 0.7262 & 0.2793 & 91.4 & 13.9 \\

\hline
\end{tabular}
\caption{Performance of adaptive trace budgeting and selective refinement on Gemma-2-9B and Qwen-2.5-7B. }
\label{tab:cross_model_results}
\end{table*}
}

\subsubsection{Effect of Selective Refinement}

We next evaluate the proposed selective execution-guided refinement strategy. During refinement, each generated query is verified by executing against the database, and failed queries are passed to the correction model (Gemma-4 in our experiments), which generates a revised query using the execution feedback. We use a maximum of two refinement iterations in all experiments. 

We first apply refinement to all predictions, regardless of their estimated difficulty. We then investigate whether refinement can be applied more selectively using the estimated difficulty tiers introduced in Section~\ref{sec:difficulty_tier}. Specifically, we evaluate four refinement policies:
(i) \textbf{All}, where all predictions are verified and failed generations are refined;
(ii) \textbf{Empty}, where refinement is applied only to predictions rejected by grammar/schema filtering;
(iii) \textbf{Empty+Hard}, where refinement is applied to empty predictions and hard-difficulty questions; and
(iv) \textbf{Empty+Medium+Hard}, where refinement is applied to empty predictions together with medium- and hard-difficulty questions. 

Table~\ref{tab:eval_results} summarizes the results. Applying refinement improves both lexical and execution-based performance for the fixed-budget and adaptive-budget pipelines. For the fixed-budget pipeline, refinement increases the execution success rate from 84.0\% to 92.6\%, while the adaptive-budget pipeline improves from 83.1\% to 92.4\%. Similar improvements are observed in both lexical and execution ROUGE-L scores. Paired bootstrap resampling (10,000 resamples) shows that the full refinement configurations significantly outperform their corresponding non-refined baselines. 
The additional refinement overhead is relatively small, increasing the average inference time by approximately 1.1\,s per question. This limited increase occurs because many generated queries are already executable and require only the initial verification step without additional correction iterations. 

We further analyze whether refinement can be applied selectively. Refining empty predictions together with medium- and hard-difficulty questions achieves nearly the same execution success rate (91.9\%) as refining all predictions (92.4\%), while reducing the number of refinement calls. In contrast, refining only empty predictions or empty predictions together with hard questions results in larger decreases in execution success (87.8\% and 88.3\%), showing the importance of refining medium-difficulty questions. 
This behavior is consistent with the distribution of refinement candidates. Only 66 of 789 predictions (8.4\%) fail initial execution, of which 42 are successfully corrected. Medium-difficulty questions account for most successful corrections (about 76\%), matching their 70\% share of the evaluation set. Easy questions rarely fail (4.9\%) and contribute only four successful corrections, so excluding them from refinement reduces execution success by just 0.5\%.

These results suggest that difficulty estimation provides a useful signal for allocating refinement effort. Easy questions rarely require correction, and excluding them from refinement has only a minimal impact on execution success, while refining medium-difficulty questions is important for preserving the overall gains from execution-guided refinement. Selectively applying refinement based on predicted difficulty therefore preserves most of the benefits of execution feedback while reducing unnecessary refinement attempts.

\subsubsection{Generalization Across Models}

We evaluate whether the proposed approach transfers across generator models by repeating the experiments using Qwen-2.5-7B while keeping the same correction model (Gemma-4) used in the Gemma-2-9B experiments. This evaluates whether the proposed adaptive trace budgeting and selective execution-guided refinement strategies remain effective across generator families.

Table~\ref{tab:cross_model_results} summarizes the results. For Qwen-2.5-7B, the adaptive pipeline achieves comparable performance to the fixed-budget pipeline after applying refinement. The execution success rate changes from 92.1\% to 91.6\%, while the average inference time decreases from 17.7\,s to 14.0\,s per question, corresponding to an approximately 21\% reduction. 
Similar to the Gemma-2-9B results, selective refinement can be applied with minimal performance degradation. Applying refinement to empty predictions together with medium- and hard-difficulty questions achieves 91.4\% execution success, only 0.2\% lower than refining all predictions. This indicates that easy questions are less likely to require refinement, as skipping refinement has only a minimal impact on execution success.

An additional observation is that the same correction model, Gemma-4, remains effective for both generator models. While Gemma-4 belongs to the same model family as Gemma-2-9B, it differs from Qwen-2.5-7B. Nevertheless, execution-guided refinement improves performance in both settings, suggesting that effective refinement is possible even when the correction and generator models belong to different model families.


\section{Conclusion}\label{sec:conclusion}

Test-time inference strategies have improved the reliability of LLM-based query generation without requiring additional model training. In Text2Cypher, previous studies have explored confidence-based candidate selection and grammar-/schema-aware filtering to improve generation quality. However, these approaches generally apply uniform inference policies across questions with different levels of complexity, resulting in unnecessary computation for simple cases. Moreover, when strict filtering rejects all candidates, additional refinement may be required to recover valid queries, motivating more selective use of inference-time computation.

In this work, we introduced adaptive test-time inference strategies that dynamically allocate computation according to the expected difficulty of the input question. Adaptive trace budgeting adjusts the candidate generation budget based on estimated question difficulty, while selective execution-guided refinement applies correction only when additional inference is likely to provide benefit. Experiments with Gemma-2-9B and Qwen-2.5-7B show that adaptive trace budgeting reduces the average number of generated traces by 30.7\% and wall-clock inference time by 21--25\% while maintaining comparable generation quality. Furthermore, selective refinement achieves nearly the same improvement as refining all queries, with only a 0.2--0.5\% reduction in execution success while avoiding unnecessary refinement for low-complexity questions. Finally, using the same Gemma-4 corrector across both generator families suggests that effective refinement does not require the generator and corrector to belong to the same model family. 

Future work includes replacing the heuristic difficulty estimator with confidence signals derived directly from the generation process and extending adaptive test-time inference strategies to other query generation tasks.

\section*{Limitations}\label{sec:limitations}

Our work has several limitations. First, adaptive trace budgeting relies on heuristic difficulty estimation based on pre-generation linguistic features. Although effective on the evaluated benchmark, these heuristics may require adaptation for other Text2Query tasks. The difficulty thresholds and trace budgets were also selected heuristically based on the difficulty score distribution rather than tuned using execution or quality metrics. Future work could investigate learned difficulty estimators, automatically calibrated inference budgets, and confidence signals derived directly from the generation process. It would be useful to compare adaptive budgeting with additional fixed-budget baselines using the same average inference budget to better isolate the effect of difficulty-aware allocation.

Our evaluation also has several limitations. We evaluate the proposed approach only on Text2Cypher generation using a single benchmark. Although the proposed adaptive budgeting and selective refinement strategies are intended to be independent of a particular model or task, additional experiments on tasks such as Text2SQL and Text2SPARQL are needed to evaluate their broader applicability. We also evaluate only two generator models (Gemma-2-9B and Qwen-2.5-7B) together with a single refinement model (Gemma-4). Although the results suggest that refinement can transfer across generator families, experiments with additional open-weight and proprietary models would provide stronger evidence. Finally, inference times were measured under our experimental hardware configuration, so absolute latency may differ across deployment environments. However, the relative computational savings of adaptive inference are expected to remain similar.


\bibliography{main}

\appendix

\section{Declaration of AI assistance}
AI-based writing and coding assistants were used during the preparation of this work to improve the wording and clarity of the manuscript, assist with auxiliary analysis and evaluation scripts, and discuss implementation options. All AI-assisted text and code were reviewed, edited, and verified by the author(s). The research and reported results were conducted and verified by the author(s). AI assistants were not used to generate experimental results or references.


\end{document}